%Paper: hep-th/9509003
%From: soldati@bo.infn.it
%Date: Fri, 1 Sep 1995 17:20:22 +0200

\magnification=1200
\hsize=15 truecm
\vsize=22 truecm
\baselineskip 20 truept
\voffset=1.5 truecm
\parindent=1cm
\overfullrule=0pt

  %\input nuovisim.tex
  %\input automacro.tex
  %\autobibliografia
  %\stileincite[]\stileinbibl[]
  %\numerazionedoppia
  %\riferimentifuturi

\def\vsp{\vskip 0.4truecm \par}
\def\ts{\thinspace}

\def\ni{\noindent}

\def\rp{\item{}}
\def \r#1{[\ts {#1}\ts \ts ]} \def\rf#1{\item{\r{#1} \ }}

\centerline{\bf ON THE SCATTERING AMPLITUDE }
\centerline{\bf IN THE AHARONOV--BOHM GAUGE FIELD }

\vskip 1truecm

\centerline{{\bf  Paola Giacconi $^{(\ast)}$}
\footnote{$^{(1)}$}{{\rm E--mail :
giacconi, soldati@infn.bologna.it} }, {\bf Fabio Maltoni $^{(\flat)}$}
\footnote{$^{(2)}$}{{\rm E--mail : maltoni@ipifidpt.difi.unipi.it}},
{\bf Roberto  Soldati $^{(\sharp)}$} }

\vskip 1truecm

\centerline{ $^{(\ast)}$ Dipartimento di Matematica e Fisica,}
\centerline{Camerino, Italia}
\centerline{- sezione I.N.F.N. Bologna, Italia }
\centerline{$^{(\flat)}$ Dipartimento di Fisica, Pisa}
\centerline{- sezione I.N.F.N. Pisa, Italia}
\centerline{$^{(\sharp)}$ Dipartimento di Fisica "A. Righi", Bologna}
\centerline{- sezione I.N.F.N. Bologna, Italia }

\vskip 2truecm

\centerline{ABSTRACT}
\ni
A general expression for the scattering amplitude of nonrelativistic
spinless particles in the Aharonov-Bohm gauge potential is obtained
within the time independent formalism. The result is valid also in
the backward and forward directions as well as for any choice of the
boundary conditions on the wave function at the flux tube position.

\vskip 3.0 truecm
{\bf DFUB/95 - 6}\par
September 1995
\vfill\eject
\noindent
{\bf 1.\quad Introduction}
\medskip
Since the appearence of the seminal paper [1] concerning the scattering
on the plane of spinless charged particles in the field of an infinitely
thin flux tube, the so called Aharonov-Bohm effect and related topics have
been the sources of a very large number of investigations. Notwithstanding
the really wide Literature on the subject, there are some few but relevant
points, even within the originally considered scattering problem, which
still  deserve a more general mathematical
formulation and, consequently, a more transparent physical interpretation,
at least in our opinion.
By the way the Aharonov-Bohm problem has recently gained a revival of
interest, mainly because of its application to the anyon matter [2],
to the cosmic strings [3] and to the planar gravity [4].\par
Basically there are two issues which still need a further analysis.
The first one deals with the evaluation of the scattering amplitude in the
backward and forward directions. To our knowledge this point has been
firstly discussed in a very clear and rigorous way in ref. [5]. In this paper
it is shown that, strictly speaking, the celebrated formula for the
differential cross section holds true outside two opposite narrow
cones, with common vertices at the point-like flux tube position, which
contain the backward and forward directions respectively. As a matter of fact,
on the one hand the asymptotic formulae, which allow to single out the
scattering amplitude, are valid outside the above mentioned narrow cones.
On the other hand, the alternative method described in ref. [5]
to treat the problem, which
makes use of a contour integral representation of the Bessel functions,
is such that the separation between incident and scattered waves is no
longer possible in the backward and forward directions; {\it sic
stantibus rebus} a meaningful definition and eventual evaluation of the
backward and forward scattering amplitudes are still lacking. There have
been proposals [4],[6] to introduce some regularizations, in order to
give a meaning to the sum over the partial phase shifts, which turns out
to be meaningless as it stands. Then, however, delta-like singularities
in the aforementioned critical directions unavoidably appear. As far as
we understand, those singularities turn out to be awkward from the
mathematical point of view [7]
and, more important, they result to be a genuine
effect of some specific choice (presumably not the most convenient) of
a regularization procedure; consequently, it seems that their
presence could likely be avoided.\par
The second issue is a concern of the choice of the boundary conditions
for the wave function at
the position of the infinitely thin flux tube. In the standard treatment
the wave function is chosen to vanish at the flux tube position: this
is a possible and simple choice, but far from being the most general one.
Actually, a two real parameters family of boundary conditions exists,
which stays into one-to-one correspondence with the self-adjoint extensions
of the quantum Hamiltonian operator [4],[8]. This fact entails that a
more general formula for the differential cross section has to be considered,
which generalizes the original Aharonov-Bohm famous result and
reproduces the latter one for a special choice of the parameters of the
self-adjoint extensions. As a by-product one can also get a clear
understanding of the limits and the meaning of the Born approximation.\par
In this work we shall obtain a quite general formula for the scattering
amplitude and the differential cross section, which includes the backward
and forward directions as well as any choice of the boundary
conditions on the wave function. The basic ingredients we shall develop here
are the concept of the adiabatic switching of the interaction and the method
of the analytic continuation of the scattering amplitudes.
The paper is organized as follows. In sect. 2 we discuss the general
framework and derive the formula for the scattering amplitude arising from
the regular part of the wave function. In sect. 3 we treat the subspaces
of the Hilbert space which are influenced by the choice of the boundary
conditions at the flux tube position or, equivalently, by the self-adjoint
extensions. In sect. 4 we discuss some consequences of our treatment, while
the technical details leading to the main formulae are presented in the
appendix.
\bigskip
\noindent
{\bf 2.\quad The scattering amplitude without the $S$- and $P$-waves}
\medskip
The standard non relativistic Hamiltonian for a spinless charged particle,
with charge $e$ and mass $m$,
moving on the plane in the presence of some infinitely thin flux tube
located at the origin is given by
$$
H={1\over 2m}\left[{\bf p} - e{\bf A}({\bf r})\right]^2\ ,
\eqno(2.1)
$$
where the Aharonov--Bohm (AB) gauge potentials is described as
$$
A_j({\bf r})={\alpha\over e}\epsilon_{jk}{x_k\over r^2}\quad j,k=1,2\ ,
\eqno(2.2)
$$
with $-1<\alpha<0$, which is the range we are here interested in.
Setting $\epsilon_{12}=1$, the field strength can be written as
$$
F_{12}({\bf r})={2\pi|\alpha|\over e}\delta^{(2)}({\bf r})\ .
\eqno(2.3)
$$
If we split the Hamiltonian into free and interaction parts
$$
H({\bf p},{\bf r})={{\bf p}^2\over 2m}+V({\bf p},{\bf r})\ ,
\eqno(2.4)
$$
we obtain, going into polar coordinates $(r,\varphi)$,
$$
V\left(r,{\partial\over \partial\varphi}\right)={\alpha\over 2mr^2}
\left(-2i{\partial\over \partial\varphi}+\alpha\right)\ .
\eqno(2.5)
$$
The solutions of the eigenvalue problem $H\psi_k^{AB}=E\psi_k^{AB}$,
with $E=k^2/2m,\ k\ge 0$, corresponding to
a particular set of stationary scattering states,
are given by
$$
\psi_k^{AB}(r,\varphi)=(2\pi)^{-1/2}\sum_{n=-\infty}^{+\infty}
J_{|n+\alpha|}(kr)\exp\left\{in\varphi-i{\pi\over 2}|n+\alpha|\right\}\ .
\eqno(2.6)
$$
These eigenfunctions fulfil the following properties [1],[5]: namely,
$$
\eqalignno{
 i & )\quad  \psi_k^{AB}(r=0,\varphi)=0\ ; & (2.7a) \cr
ii & )\quad  \psi_k^{AB}(r,\varphi)\ \sim\
\phi_k^{AB}(r,\varphi)+f^{AB}(k,\varphi)
{e^{ikr}\over \sqrt{2\pi r}}\quad ({\rm large}\ r)\ ; & (2.7b) \cr
iii &)\quad  \langle\psi_k^{AB}|\psi_p^{AB}\rangle\ = \
{1\over k}\delta (k-p)\ ; & (2.7c) \cr}
$$
here the wave functions $\phi_k^{AB}(r,\varphi)$ do not correspond to
eigenfunctions
of the free Hamiltonian $H_0={\bf p}^2/2m$, owing to the long range nature of
the AB potential; they are given instead by the
phase factors
$$
\phi_k^{AB}(r,\varphi)={1\over \sqrt{2\pi}}
\exp\{-ikr\cos\varphi-i\alpha\varphi\}\ .
\eqno(2.8)
$$
It is worthwhile to stress that
the asymptotic behaviour ii) holds true outside the
sufficiently narrow cones $|\varphi|<\pi-{\cal O}[(kr)^{-1/2}]$
and, consequently,
the phase factor $\phi_k^{AB}$ turns out to be single valued. The asymptotic
behaviour of eq.~(2.7b) can be obtained either
from an integral representation
[1],[9] for the exact solution of eq.~(2.6), or from a  contour integral
representation of the Bessel functions [4],[5].\par
Since the eigenfunctions $\psi_k^{AB}$ are not normalizable, the scalar
product in iii) is understood in the improper sense, $i.e.$ in the tempered
distributions topology
$$
\eqalign{
& \left<\psi_k^{AB}|\psi_p^{AB}\right>\ = \cr
& {\cal S}^\prime-\lim_{R\to\infty}\int_0^R rdr\int_0^{2\pi}d\varphi\
[\psi_k^{AB}(r,\varphi)]^*\psi_p^{AB}(r,\varphi)\ . \cr}
\eqno(2.9)
$$
\par
It turns out that the set $\{\tilde\psi_k^{AB}|
\tilde\psi_k^{AB}(r,\varphi)=\sqrt k\psi_k^{AB}(r,\varphi),\ k\ge 0\}$
is a  complete orthonormal set of improper
scattering states of the Hilbert space
${\cal H}$. As a matter of fact the Hamiltonian (2.1), together with the
boundary condition i), is a self-adjoint operator which does not admit bound
states.\par
Now the main point. In order to define the scattering amplitude, there
basically are two attitudes in the Literature. In the first one [1],[5] the
amplitude is obtained from the asymptotic behaviour of the
exact regular scattering solution of eq.~(2.6).
As already noticed, this point of view gives
rise to the sum of a
phase factor $\phi_k^{AB}= (2\pi)^{-1/2}\exp\{-i(kr\cos\varphi+\alpha
\varphi)\}$ and a scattered wave $\psi_{sc}=(2\pi r)^{-1/2}f^{AB}(k,\varphi)
\exp\{ikr\}$,
the amplitude corresponding to the well known
AB scattering amplitude [1]. However, as already  noticed,
the above mentioned asymptotic form  holds true for $\varphi\not=
l\pi,\ l\in {\bf Z}$; in other words, the above decomposition of the exact
wave function, in the large-$r$ limit, strictly speaking loses its meaning
in the backward and forward directions [5].\par
The second approach \footnote{$^{(*)}$}{To be specific, the authors of
ref.s [4] actually treat the scattering on a spinning cone; nonetheless,
the mathematical framework is in close correspondence to the case under
consideration.}[4],[6]
 attempts to obtain the scattering amplitude as
the sum of the partial amplitudes: namely,
$$
\eqalign{
f(k,\varphi)
& ={1\over \sqrt{2\pi ik}}\sum_{n=-\infty}^{+\infty}
\left(e^{2i\delta_n}-1\right)e^{in\varphi}\cr
& =
\lim_{\epsilon\to 0^+}{1\over \sqrt{2\pi ik}}\sum_{n=-\infty}^{+\infty}
\left(e^{2i\delta_n}-1\right)e^{in\varphi-|n|\epsilon}\cr
& =  f_{reg}(k,\varphi)\ .\cr}
\eqno(2.10)
$$
Within this approach the regular solutions of the Schr\"odinger equation
 are  written in the form
$$
\psi_k(r,\varphi)=\sum_{n=-\infty}^{+\infty} e^{i\delta_n+n\pi/2}
u_n(kr)e^{in\varphi}\ ,
\eqno(2.11)
$$
whose asymptotic behaviours are provided by the usual expression
$$
\psi_k(r,\varphi)\ \sim\
{1\over \sqrt{2\pi}}e^{ikr\cos\varphi}+
f_{reg}(k,\varphi){e^{ikr}\over \sqrt{2\pi r}}\quad ({\rm large}\ r)\  ,
\eqno(2.12)
$$
where $f_{reg}(k,\varphi)$ is formally given by the regularized quantity of
eq.~(2.10).\par
In so doing, however, the regularized amplitude appears to involve a
mathematically poorly defined series of
angular delta distributions in the forward and backward directions, also
leading to troubles in the definition of the differential
scattering cross section. Moreover, the remaining part of the amplitude
actually reproduces the AB expression that is known to be correct only for
$|\varphi|<\pi$. The authors of ref.s [4] suggest to move, in some sense,
 the delta-like
contribution from the scattered wave to the incident plane wave
in the limit of
large-$r$, namely to redefine the decomposition of the wave function in
the large-$r$ asymptote. Nevertheless, in order to implement this alternative
decomposition,
they eventually make use of some contour representation for the wave
function and, therefore, they resort indeed to the first approach previously
discussed. In conclusion, it seems that those angular delta distributions
appear somewhat to be an artefact of the regularization of eq.~(2.10),
while the interpretation of the amplitude in the forward and backward
directions remains admittedly still not quite clear [7].\par
The method we here develop is based on the standard integral equation
satisfied by the scattering wave functions $\psi_k^{(+)}(r,\varphi)$.
To this aim, it is convenient to split the wave function into the sum
of the $S$-wave, carrying zero angular momentum,
the $P$-wave of angular momentum $n=1$ and the rest: namely,
$$
\psi_k^{(+)}(r,\varphi)\equiv \chi_{0,k}^{(+)}(r)+
\chi_{1,k}^{(+)}(r,\varphi)+\Psi_k^{(+)}(r,\varphi)\ ,
\eqno(2.13)
$$
where
$$
\Psi_k^{(+)}(r,\varphi)\ =
\sum_{n=2}^\infty {i^n\over \sqrt{2\pi}}e^{in\varphi-i\pi\alpha/2}
J_{n+\alpha}(kr)+\sum_{n=1}^\infty {i^n\over \sqrt{2\pi}}
e^{-in\varphi+i\pi\alpha/2}J_{n-\alpha}(kr)\ .
\eqno(2.14)
$$
The $S$- and $P$-waves will be discussed separately
in the next section; the regular part
$\Psi_k^{(+)}(r,\varphi)$ of the full eigenstate,
which is different from the corresponding part in
eq.~(2.6), is constructed in such a way that the $n$-th partial wave has the
usual large-$r$ asymptote, $i.e.$
$$
\eqalign{
\int_0^{2\pi}{d\varphi\over \sqrt{2\pi}}\ & e^{-in\varphi}\left[
\Psi_k^{(+)}(r,\varphi)-{1\over \sqrt{2\pi}}e^{ikr\cos\varphi}\right] \cr
\sim & {e^{ikr}\over \sqrt{2\pi ikr}}(e^{-i\pi\alpha}-1)\quad
({\rm large}\ r)\ ,\cr}
\eqno(2.15a)
$$
from the first sum in RHS of eq.~(2.14) while
$$
\eqalign{
\int_0^{2\pi}{d\varphi\over \sqrt{2\pi}}\ & e^{in\varphi}\left[
\Psi_k^{(+)}(r,\varphi)-{1\over \sqrt{2\pi}}e^{ikr\cos\varphi}\right] \cr
\sim & {e^{ikr}\over \sqrt{2\pi ikr}}(e^{i\pi\alpha}-1)\quad
({\rm large}\ r)\ ,\cr}
\eqno(2.15b)
$$
from the second sum in the RHS of eq~(2.14).
 By its very construction, the wave function
$\Psi_k^{(+)}({\bf r})$ satisfies the scattering integral equation
$$
\Psi_k^{(+)}({\bf r})-\Phi_k({\bf r})=
{m\over i}\int d^2{\bf r}^\prime\ H_0^{(1)}(k|{\bf r}-{\bf r}^\prime|)
V({\bf p}^\prime,{\bf r}^\prime)\Psi_k^{(+)}({\bf r}^\prime)\ ,
\eqno(2.16)
$$
where $H_0^{(1)}$ is the first Hankel's function and
$$
\Phi_k({\bf r})=(2\pi)^{-1/2}\left[e^{ikr\cos\varphi}-J_0(kr)
-i e^{i\varphi}J_1(kr)\right]\ .
\eqno(2.17)
$$
\par
In order to extract from eq.~(2.16) an expression for the related
scattering amplitude, it is useful to introduce the concept of the
adiabatic switching of the interaction, as it is customary in the
perturbative field theory [10]. To this aim we consider, for instance,
a smooth function
$g_{\rho,R}(|{\bf r}|)$, with $\rho<R$ and $0\le g_{\rho,R}(r)\le 1$,
such that $g_{\rho,R}(r)=0,\ r\ge R$ and
$g_{\rho,R}(r)=1,\ 0\le r\le\rho$, which represents the extent of switching
on the interaction within a disk of radius $R$.
In so doing, we can rewrite the RHS of eq.~(2.16) in the form
$$
\eqalign{
& \Psi_k^{(+)}({\bf r})-\Phi_k({\bf r})=\cr
& \lim_{\rho,R\to\infty}{m\over i}\int d^2{\bf r}^\prime\
g_{\rho,R}(r^\prime)H_0^{(1)}(k|{\bf r}-{\bf r}^\prime|)
V({\bf p}^\prime,{\bf r}^\prime)\Psi_k^{(+)}({\bf r}^\prime)\ ,\cr}
\eqno(2.18)
$$
the limit being understood in the ${\cal S}^\prime$-topology and taken
at the very end. Now, as long as $\rho$ and $R$ are fixed, we can safely
use the asymptotic formula for the Hankel's function, when $r$ is sufficiently
large, leading to the result
$$
\eqalign{
& \Psi_k^{(+)}({\bf r})-\Phi_k({\bf r})\ \sim\
{-im\over \sqrt{2\pi ik}}
{e^{ikr}\over \sqrt{r}}\ \times\cr
& \lim_{\rho,R\to\infty}\
\int d^2{\bf r}^\prime\
g_{\rho,R}(r^\prime)\exp\{-i{\bf k}\cdot {\bf r}^\prime\}
V({\bf p}^\prime,{\bf r}^\prime)\Psi_k^{(+)}({\bf r}^\prime)\quad
({\rm large}\ r)\ , \cr}
\eqno(2.19)
$$
where ${\bf k}\equiv k{\bf r}/r$ is the momentum of the incident plane wave.
Taking eventually the limits $\rho,R\to\infty$, we finally obtain a definition
of the scattering amplitude, apart from the $S$- and $P$- waves, given by
$$
\eqalign{
F^{(+)}_k(\varphi)
& = {2\pi\alpha \over i \sqrt{2\pi ik}}\
\upsilon(\varphi,\alpha)\times \cr
& \sum_{n=2}^\infty\ \int_0^\infty {dx\over x}\ \left[
e^{-in\varphi-i\pi\alpha/2}\
J_{n+\alpha}(x)J_n(x)\ +
e^{in\varphi+i\pi\alpha/2}\ J_{n-\alpha}(x)J_n(x)\right] \cr
& + {2\pi\alpha \over i \sqrt{2\pi ik}}\
\upsilon(\varphi,\alpha)\int_0^\infty {dx\over x}\
e^{i\varphi+i\pi\alpha/2}\ J_{1-\alpha}(x)J_1(x)\ ,
\cr}
\eqno(2.20)
$$
where $\upsilon(\varphi,\alpha)\equiv \alpha+2i(\partial/\partial\varphi)$
and $x\equiv kr^\prime$.
It should be gathered that the interaction can be fully switched on
({\it i.e.} $\rho,R\to\infty$), provided the order of the operations is
fixed as in eq.~(2.20): that means, first radial integration, then sum over
angular momenta and finally derivative with respect to the angle.
In so doing, the amplitude is perfectly well defined, without need
of any {\it ad hoc} regularization.
We remark that, in the case of short-range interactions,
the adiabatic limit in eq.~(2.19) is trivial and the standard textbook
formula is immediately recovered.
Moreover, the above definition (2.19) is equivalent to the one
which is obtained from the usual asymptotic behaviour
of the scattering wave eigenfunctions whereas, for a long-range potential,
this is not exactly true in general, as the present analysis indeed shows.

In the specific AB scattering problem eq.~(2.20) entails ($l\in {\bf Z}$)
$$
 F_k^{(+)}(\varphi)=
{1\over \sqrt{2\pi ik}}\left[1-\cos\pi\alpha-
e^{-i\varphi}\left(e^{-i\pi\alpha}-1\right)-\sin\pi\alpha
\cot{\varphi\over 2}\right]\ ,
\eqno(2.21)
$$
when $\varphi\not=2l\pi$. Furthermore, the following expression holds true
in the forward direction: namely,
$$
F_k^{(+)}(\varphi=2l\pi)
={1\over \sqrt{2\pi ik}}[2(1-\cos\pi\alpha)+
i(1+\alpha)\sin\pi\alpha]\ .
\eqno(2.22)
$$
The details are given in the appendix; here we would like to add some remarks.
Firstly we see that eq.~(2.21) is nothing but the AB amplitude, up to the
$S$- and $P$- wave contributions; furthermore, from
eq.~(2.19), it is apparent that the incoming wave is a standard plane wave
and no angular delta distributions ever appear. This in turn entails that,
at variance with the formula
(2.7b), the amplitude can be singled out and evaluated,
by means of a unique analytic
continuation, also in the backward and forward directions.
Finally, we notice that eq.~(2.22) could also be reinterpreted
in the framework of the time dependent approach,
as proposed in ref. [9].
\bigskip
\noindent
{\bf 3.\quad The $S$- and $P$-wave scattering amplitudes}
\medskip
In order to properly treat the $S$- and $P$-waves,
we have to consider two facts.
The first one is to recognize that, in general, regularity at $r=0$ is
too a strong requirement for the eigenfunctions of zero
and one angular momentum.
As a matter of fact, self-adjointness of the Schr\"odinger Hamiltonian in
eq.~(2.1) can be indeed fulfilled even by singular $S$- and $P$-wave functions
(although locally square integrable around the origin) [8].
Consequently, the vanishing boundary condition at the origin as in
eq.~(2.7a) does not appear to be a necessary requirement, for it has
to be suitably generalized.
The second one
is that, owing to the presence of a singular part in the above mentioned
partial waves,
the corresponding scattering amplitude has to be necessarily obtained from
the large-$r$ asymptote of some $S$- and $P$-waves satisfying the most
general boundary
conditions we shall specify below; actually, a definition of those
partial waves amplitudes following the method of the
 previous section is no longer available, due
to the infrared divergencies in the integral of eq.~(2.20).
According to the Von Neumann theory of the deficiency indices [11], one
obtains the most general behaviour at the origin of the wave function:
namely,
$$
\psi({\bf r})\ \sim\ \sum_{n=0,1} C_n r^{\gamma (n)}\quad ({\rm small}\ r)\ ,
\eqno(3.1)
$$
where $C_n$ are some $r$-independent quantities and $-1<\gamma (n)<0$.

With the boundary condition established in eq.~(3.1) the $S$- and $P$-
scattering wave functions become
$$
\eqalignno{
\chi_{0,k}^{(+)}(r)
& = {\exp(i\pi\alpha/2)\over \sqrt{2\pi}
(1+i\tan\pi\mu_0)}
\left[J_{-\alpha}(kr)+\tan\pi\mu_0 N_{-\alpha}(kr) \right]\ ,
& (3.2a)\cr
\chi_{1,k}^{(+)}(r,\varphi)=
& -
{\exp[-i(\pi/2)(1+\alpha)+i\varphi]\over
\sqrt{2\pi}(1+i\tan\pi\mu_1)}\cr
& \times \left[J_{1+\alpha}(kr)+\tan\pi\mu_1 N_{1+\alpha}(kr) \right]\ ,
& (3.2b)\cr}
$$
where
$$
\eqalignno{
&\tan\pi\mu_0(k)\equiv {\sin\pi\alpha\over\cos\pi\alpha
-(k^2/2mE_0)^\alpha}\ , &(3.3a) \cr
&\tan\pi\mu_1(k)\equiv {\sin(-\pi\alpha)\over\cos\pi\alpha
+(k^2/2mE_1)^{-(1+\alpha)}}\ , &(3.3b) \cr}
$$
$E_0\ ,\ E_1$ being bound state energies for the angular momenta $n=0$ and
$n=1$ respectively. Actually, for any choice of the self-adjoint extensions
corresponding to fixed values of the parameters $\mu_0\in {\bf R},\ \mu_1\in
{\bf R}$,
there exist solutions of the equations
$H B_0(r)=-E_0B_0(r)$, for $n=0$ and $E_0>0$, $H B_1(r,\varphi)
=-E_1B_1(r,\varphi)$, for $n=1$ and $E_1>0$. The normalized
eigenfunctions are
$$
\eqalign{
B_0(r) & = (\kappa_0/\pi)\sqrt{\sin(-\pi\alpha)}\ K_\alpha(\kappa_0r)\ ,\cr
B_1(r,\varphi) & = (\kappa_1/\pi)\sqrt{\sin\pi(1+\alpha)}\
K_{1+\alpha}(\kappa_1r)e^{i\varphi}\ ,\cr
\kappa_{0(1)} & \equiv \sqrt{2mE_{0(1)}}\ ,\cr}
\eqno(3.4)
$$
which turn out to be orthogonal to the scattering eigenfunctions.
In order to obtain  regular
scattering wave functions from eq.s~(3.2) one has to take
the limits $E_0\to\infty$ and $E_1\to\infty$ that
give $\tan\pi\mu_0=\tan\pi\mu_1=0$. It has to be gathered that,
as already noticed, in the above limits the bound states disappear as
the corresponding wave functions (3.4) indeed identically vanish.
Accordingly, the regular scattering wave functions are a complete
orthonormal set which diagonalize the self-adjoint invertible quantum
Hamiltonian and, safely, there do not appear bound states with infinitely
negative energies, whose physical interpretation turned out to be
troublesome.

On the other hand,
taking the limits $E_0\to 0$ and $E_1\to 0$, from which
$\tan\pi\mu_0=-\tan\pi\mu_1= \tan\pi\alpha$, one is led to the purely
singular eigenfunctions.
Moreover, in the absence of AB potential $(\alpha \to 0)$, eq.s~(3.2) give
scattering wave functions in the presence of a purely contact interaction;
we have to stress that, in this limit, only the $S$- wave exhibits
a singular part: namely,
$$
\eqalignno{
&\lim_{\alpha\to 0}\tan\pi\mu_0 =-{\pi \over \ln(k^2/2mE_0)}\ ,&(3.5a)\cr
&\lim_{\alpha\to 0}\tan\pi\mu_1=0\ .&(3.5b)\cr}
$$
In other words we can say that the $P$- wave is not influenced by the
pure $\delta$-like potential.
Now we are ready to compute the scattering amplitudes for $S$- and $P$-
waves from their large-$r$  behaviour. They read
$$
\eqalignno{
& f^{(+)}_{0}(k;E_0)={1\over \sqrt{2i\pi k}}\left[e^{i\pi\alpha}
(\cos 2\pi\mu_0 - i\sin 2\pi\mu_0)-1\right]\ , &(3.6a)\cr
& f^{(+)}_{1}(k,\varphi;E_1)={e^{-i\varphi}\over \sqrt{2i\pi k}}
\left[e^{-i\pi\alpha}(\cos 2\pi\mu_1 - i\sin 2\pi\mu_1)-1\right] \quad .
&(3.6b)\cr}
$$
It should be noticed that, from eq.s (3.6), a nonvanishing contribution to
the scattering amplitude arises also in the backward and
forward directions. It is also interesting to consider
the limit $\alpha \to 0$  in eq.s~(3.6): namely,
$$
\eqalign{
& \lim_{\alpha\to0}f^{(+)}_{0}(k;E_0)=
 {1\over \sqrt{2i\pi k}}
\left[\exp\left\{2i\arctan{\pi\over \ln(2mE_0/k^2)}\right\}-1\right]\ , \cr
& \lim_{\alpha\to0}f^{(+)}_{1}(k,\varphi;E_1)=0\ , \cr}
\eqno(3.7)
$$
which confirms that the $P$- wave is not influenced
by a $\delta$-like potential.\par
In order to obtain the usual AB scattering amplitude we have to consider
the (regular) limits $E_0\to\infty$ and $E_1\to\infty$ of eq.s~ (3.6);
this entails
$$
\eqalignno{
&\lim_{E_0\to\infty} f^{(+)}_{0}(k;E_0)={1\over \sqrt{2i\pi k}}
(e^{i\pi\alpha}-1)\ , &(3.8a)\cr
&\lim_{E_1\to\infty} f^{(+)}_{1}(k,\varphi;E_1)=
{e^{-i\varphi}\over \sqrt{2i\pi k}} (e^{-i\pi\alpha}-1)\ ; &(3.8b)\cr}
$$
by adding both the above expression to eq.~(2.21a), one easily reconstruct
the well known AB formula.\par
It is also easy to recover the phase shifts corresponding to the AB
and purely contact $\delta$-like potentials ($\alpha \to 0$) for angular
momenta $n=0$ and $n=1$  respectively: they read,
$$
\eqalignno{
& \delta_0(k,\alpha;E_0)=\cr
& {\pi\alpha\over 2}-\arctan{\sin\pi\alpha\over \cos\pi\alpha -
(k^2/2mE_0)^\alpha}
\buildrel \alpha\to 0 \over \longrightarrow
\arctan {\pi\over\ln(2mE_0/k^2)}\ ; &(3.9a)\cr
& \delta_1(k,\alpha;E_1)=
-{\pi\alpha\over 2}+\arctan{\sin\pi\alpha\over \cos\pi\alpha -
(k^2/2mE_1)^{-(\alpha+1)}}\ . &(3.9b)\cr}
$$
It should be noticed that eq.s~(3.9) correctly reproduce all the limits
$\alpha\to 0$ and $E_0, E_1\to\infty$, at variance with the incorrect formula
of the first paper in ref. [8].
\bigskip
\noindent
{\bf 4.\quad Discussion}
\medskip
In the presence of a general behaviour at the origin, as expressed by the
relation (3.1), the set $\{\tilde\psi_k^{AB}|k\ge 0\}$ no longer represents
a (improper) basis in the Hilbert space.
We have instead to select, for instance, the following
complete orthonormal family $(k\ge 0)$
$$
\eqalignno{
\chi_{k,n}(r,\varphi)
& =\sqrt{{k\over 2\pi}}J_{|n+\alpha|}(kr)\exp\{in\varphi\}
\quad n\not= 0,1\ ; &(4.1a)\cr
\chi_{k,0}(r;E_0)
& =\sqrt{{k\over 2\pi}}\left\{\cos\pi\mu_0(k)J_{|\alpha|}(kr)+
\sin\pi\mu_0(k)N_{|\alpha|}(kr)\right\}\ ; &(4.1b)\cr
\chi_{k,1}(r,\varphi;E_1)
& =\sqrt{{k\over 2\pi}}\left\{\cos\pi\mu_1(k)J_{1-|\alpha|}(kr)\right. \cr
& +\left.\sin\pi\mu_1(k)N_{1-|\alpha|}(kr)\right\}\exp\{i\varphi\}\ ;
&(4.1c)\cr
B_0(r)
& = (\kappa_0/\pi)\sqrt{\sin(-\pi\alpha)}\ K_\alpha(\kappa_0r)\ ; &(4.1d)\cr
B_1(r,\varphi)
& = (\kappa_1/\pi)\sqrt{\sin\pi(1+\alpha)}\
K_{1+\alpha}(\kappa_1r)e^{i\varphi}\ . &(4.1e)\cr}
$$
In particular, according to the fundamental theorem for self-adjoint operators,
the following completeness relation must hold: namely,
$$
\eqalign{
& \int_0^\infty {kdk\over 2\pi}\left\{\cos^2\pi\mu_0(k)J_{|\alpha|}(kr)
J_{|\alpha|}(kr^\prime) + \sin^2\pi\mu_0(k)N_{|\alpha|}(kr)
N_{|\alpha|}(kr^\prime)\right.\cr
& + \left.(1/2)\sin2\pi\mu_0(k)\left(J_{|\alpha|}(kr)N_{|\alpha|}(kr^\prime)
+J_{|\alpha|}(kr^\prime)N_{|\alpha|}(kr)\right)\right\}
=\delta^{(2)}({\bf r}-{\bf r}^\prime)\ , \cr}
\eqno(4.2)
$$
and the analogous relation for the $n=1$ angular momentum Hilbert subspace.
The above completeness relations turn out to be quite useful,
in order to construct
integral kernels of operators commuting with the Hamiltonian; furthermore,
such a kind of resolutions of the identity become powerful tools,
for instance, in the calculation of physical
quantities such as the second virial coefficient of the anyon gas [12]
or the anomaly of the axial current
in the spinning case [13]. It should also be noticed that the normalization
of the improper scattering states in eq.s~(3.2) is, instead, rather
cumbersone in general; on the other hand they exhibit, by their very
construction, a standard asymptotic behaviour. These features are in
contrast with the properties of
the regular scattering states $\tilde\psi_k^{AB}(r,\varphi)$, which
fulfil standard normalization (see eq.~(2.7c)) but non standard
asymptotic form (see eq.~(2.7b)).\par
Another interesting aspect of the present approach is that it allows
a direct comparison with the Born approximation, as it is apparent from
eq.~(2.20). As a matter of fact, by adding the RHS of eq.~(3.8b), which
represents the contribution of the regular $P$-wave, to the amplitude in
eq.~(2.21) we easily obtain, up to the first order in the coupling $\alpha$,
the Born approximation [14], $i.e.$
$$
f_{Born}^{(+)}(k,\varphi)={|\alpha|\pi\over \sqrt{2\pi ik}}\cot{\varphi\over
2}\ .
\eqno(4.3)
$$
Actually, the above expression is fully determined by the (regular)
components of the plane waves with
non vanishing angular momentum, since the $S$-wave leads to a divergent
contribution but ${\cal O}(\alpha^2)$. This means that, within the Born
approximation, contact-like interaction can not be taken into account; as a
consequence the expansion of the full
amplitude
$$
f^{(+)}_k(\varphi;E_0,E_1)\equiv
F^{(+)}_k(\varphi)+f^{(+)}_0(k,\varphi;E_0)+f^{(+)}_1(k,\varphi;E_1)\ ,
\eqno(4.4)
$$
up to the first order in $\alpha$, obviously does not lead to eq.~(4.3).\par
As a final comment, we would like to observe that the full amplitude
in eq.~(4.4) when $\varphi=2l\pi,\ l\in {\bf Z}$, gives rise to a finite
differential cross section in the forward direction. This means that,
even within the time-independent approach, the concept of the adiabatic
switching of the interaction allows a complete
proper description of the AB physics
at variance with the usual phase shifts analysis. Notice that the inadequacy
of the phase shifts approach in the treatment of the AB problem has led
the author of ref. [9] to develop the time dependent approach which,
however, appears to be more involved than the time independent one,
at least in our opinion. Furthermore, in that paper an explicit expression
is not given for the forward scattering amplitude, but only arguments
concerning the scattering of some wave packets.\par
We also notice that the present analysis might be generalized to the
spinning case as well as to the scattering of particles in a
$2+1$ dimensional space-time in the presence of gravitational fields.
\bigskip
\noindent
{\bf Acknowledgments}:
\medskip
This work has been partially supported by a grant from Ministero della
Universit\`a e della Ricerca Scientifica e Tecnologica - quota 40\%\ .
\bigskip
\noindent
{\bf Appendix}
\medskip
Here we want to show how eq.s~(2.21-22) can be derived.
The starting point is the scattering amplitude, defined by eq.~(2.20),
whose RHS we want here to evaluate it in a straightforward way.
Taking eq.s~(2.5) and (2.14) into account, after integration over
$\varphi^\prime$ and rescaling of the integration variable
$kr^\prime\equiv x$ respectively,
we obtain
$$
\eqalign{
F^{(+)}_k(\varphi)
& = {2\pi\alpha \over i \sqrt{2\pi ik}}\
\upsilon(\varphi,\alpha)\times \cr
& \sum_{n=2}^\infty\ \int_0^\infty {dx\over x}\ \left[
e^{-in\varphi-i\pi\alpha/2}\
J_{n+\alpha}(x)J_n(x)\ +
e^{in\varphi+i\pi\alpha/2}\ J_{n-\alpha}(x)J_n(x)\right] \cr
& + {2\pi\alpha \over i \sqrt{2\pi ik}}\
\upsilon(\varphi,\alpha)\int_0^\infty {dx\over x}\
e^{i\varphi+i\pi\alpha/2}\ J_{1-\alpha}(x)J_1(x)\cr
& ={2 \over i \sqrt{2\pi ik}}\sin\left({\pi\alpha\over2}\right)
\upsilon(\varphi,\alpha)
\left[\sum_{n=2}^\infty {e^{-i\pi\alpha/2-in\varphi}\over
n+(\alpha/2)} -
\sum_{n=1}^\infty {e^{i\pi\alpha/2+in\varphi}\over
n-(\alpha/2)}\right]\quad,\cr}
\eqno(A.1)
$$
where $\upsilon(\varphi,\alpha)\equiv\alpha+2i(\partial/\partial\varphi)$
and $x\equiv kr^\prime$. We want to stress that all the above expressions
are perfectly well defined, provided the operator $\upsilon(\alpha,\varphi)$ is
applied after summing the series; consequently, taking the above
specified order of operations carefully into account (integration, sum,
action of the $\upsilon$-operator), there is no need of any regularization.
As a matter of fact, from the basic formula [15]:
$$
\sum_{n=0}^\infty\ {e^{-in\varphi}\over n+z} =
\beta(z) + {e^{iz\varphi}\over2}\int_\varphi^\pi dt
e^{-izt}\left(\cot{t\over2} +i\right)\ ,
\eqno(A.2)
$$
we recover, after addition of the partial waves contributions from eq.s~(3.8),
the well known AB scattering amplitude
$$
f_{AB}^{(+)}(k,\varphi) = {1\over \sqrt{2\pi ik}}\sin\pi|\alpha|
\left(\cot{\varphi\over 2}-i\right),\ \ \varphi\not=2l\pi\ ;
\eqno(A.3)
$$
it should be enphasized that, at variance with previous derivations [1],[5],
within the present framework eq.~(A.3) holds true even in the backward
direction.
\par
Now the key point is to realize
that eq.~(A.1) allows a unique analytic continuation,
depending upon a complex variable $s$, which admits a well defined limit
at the physical value $s=1$,
even in the forward direction $\varphi=2l\pi$.
To this aim, let us introduce the analytically continued
amplitude: namely,
$$
\eqalign{
F^{(+)}_k(\varphi;s)
& \equiv{2 \over i \sqrt{2\pi ik}}
\sin\left({\pi\alpha\over 2}\right) \times\cr
& \left\{
e^{i\pi\alpha/2}\left(-{\alpha\over 2}\right)^{1-s}
-e^{-i\pi\alpha/2}\left[e^{-i\varphi}
\left(1+{\alpha\over2}\right)^{1-s}
+\left({\alpha\over 2}\right)^{1-s}\right]\right.\cr
& \left.+\sum_{n=0}^\infty\left[ {e^{-i\pi\alpha/2-in\varphi}\over
\left(n+(\alpha/2)\right)^{s-1}}
-{e^{i\pi\alpha/2+in\varphi}\over
\left(n-(\alpha/2)\right)^{s-1}}\right]\right\}\quad.\cr}
\eqno(A.4)
$$
We stress that, within the strip $Res\ge 2$, the $\upsilon$-operator
can be freely interchanged with the series since, now,
the physical limit $s\to 1$
will be performed at the very end.
In the case $\varphi\not=2l\pi \  (l \in {\bf Z})$, eq.~(A.4) becomes
$$
\eqalign{
F^{(+)}_k(\varphi;s)
& ={2i \over \sqrt{2\pi ik}}\sin\left({\pi\alpha\over 2}\right)
\left\{e^{i\pi\alpha/2}\left[\Phi(e^{i\varphi},s-1,-\alpha/2)
-\left(-{\alpha\over2}\right)^{1-s}\right]\right. \cr
& \left.-e^{-i\pi\alpha/2}\left[\Phi(e^{-i\varphi},s-1,\alpha/2)
-e^{-i\varphi}\left(1+{\alpha\over2}\right)^{1-s}
-\left({\alpha\over2}\right)^{1-s}\right]\right\}\ ;\cr}
\eqno(A.5)
$$
therefore, using the analytic extension of the function $\Phi$ [16], we get
$$
\eqalign{
\Phi(z,s=1,v)&={1\over z^v} \left[{1\over
\ln(1/z)}-\sum_{n=1}^\infty {1\over n!} B_n(v)(\ln z)^{n-1}\right]\cr
&={1\over 1-z}\ ,\qquad |\ln z|<2\pi\ ,\cr}
\eqno(A.6)
$$
where $B_n(v)$ denote the Bernoulli polynomials.
{}From the last formula we can easily recover eq.~(2.21): namely,
$$
\eqalign{
F^{(+)}_k(\varphi)
& \equiv \lim_{s\to 0^+}F^{(+)}_k(\varphi;s)
={2i\over \sqrt{2\pi ik}}
\sin\left({\pi\alpha\over 2}\right)\times\cr
& \left[e^{i\pi\alpha/2}\left({1\over 1-e^{i\varphi}}-1\right)-
e^{-i\pi\alpha/2}\left({1\over 1-e^{-i\varphi}}
-e^{-i\varphi}-1\right)\right]\cr
& ={1\over \sqrt{2\pi ik}} \left[1-\cos\pi\alpha-
e^{-i\varphi}\left(e^{-i\pi\alpha}-1\right)-\sin\pi\alpha
\cot{\varphi\over 2}\right]\ . \cr}
\eqno(A.7)
$$
Moreover, in the case $\varphi=2l\pi$,
the forward scattering amplitude of eq.~(A.4) can be defined by means of the
Hurwitz Zeta Function $\zeta(s,q)$ [16]: actually,
$$
\eqalign{
F^{(+)}_k(\varphi=2l\pi;s)
& \equiv {2i \over \sqrt{2\pi ik}}
\sin\left({\pi\alpha\over 2}\right)
\left\{e^{i\pi\alpha/2}\left[\zeta(1-s,-\alpha/2)
-\left(-{\alpha\over2}\right)^{1-s}\right]\right.\cr
& -\left. e^{-i\pi\alpha/2}\left[\zeta(1-s,\alpha/2)
-\left(1+{\alpha\over2}\right)^{1-s}
-\left({\alpha\over2}\right)^{1-s}\right]\right\}\ ,\cr}
\eqno(A.8)
$$
which easily leads to eq.~(2.22), since
$$
\eqalign{
& F^{(+)}_k(\varphi=2l\pi)
\equiv\lim_{s\to 1}F^{(+)}_k(\varphi=2l\pi;s)=\cr
& {2 \over i \sqrt{2\pi ik}}\sin\left({\pi\alpha\over 2}\right)
\left\{e^{-i\pi\alpha/2}\left[B_1(-\alpha/2)+1\right]
-e^{i\pi\alpha/2}\left[B_1(\alpha/2)+2\right]\right\}\ .\cr}
\eqno(A.9)
$$
\bigskip
\vsp
\ni{{\bf  References}} \vsp
\rf1  Y. Aharonov and D. Bohm, Phys. Rev. {\bf 115} (1959) 485.
\rf2  F. Wilczek, ``{\it Fractional Statistics and Anyon Superconductivity}'',
      World Publishing, Singapore (1990);
\rp   P. de Sousa Gerbert, Int. Jour. Mod. Phys. A {\bf 6} (1991) 173;
\rp   R. Iengo and K. Lechner, Phys. Rep. {\bf 213} (1992) 180;
\rp   S. Forte, Rev. Mod. Phys. A {\bf 7} (1992) 1025.
\rf3  A. Vilenkin, Phys. Rep. {\bf 121} (1985) 263 and references therein.
\rf4  S. Deser and R. Jackiw, Comm. Math. Phys. {\bf 118} (1988) 495;
\rp   P. de Sousa Gerbert and R. Jackiw,
      Comm. Math. Phys. {\bf 124} (1989) 229.
\rf5  M. V. Berry, R. G. Chambers, M. D. Large, C. Upstill and J. C.
      Walmsley, Eur. J. Phys. {\bf 1} (1980) 154.
\rf6  S. N. M. Ruijsenaars, Ann. of Phys. {\bf 146} (1983) 1.
\rf7  C. R. Hagen, Phys. Rev. D {\bf 41} (1990) 2015.
\rf8  C. Manuel and R. Tarrach, Phys. Lett. B {\bf 268} (1991) 222.
\rp   J. Grundberg, T. H. Hansson, A. Karlhede and J. M. Leinaas,
      Mod. Phys. Lett. B {\bf 5} (1991) 539.
\rf9  D. Stelitano, preprint MIT-CTP\# 2383 (1994).
\rf{10}  N. N. Bogoliubov and D. V. Shirkov, "{\it Introduction to the
      theory of quantized fields}",
      John Wiley \& Sons, New York (1959), p. 198.
\rf{11}  S. Albeverio, F. Gesztesy, R. Hoegh--Krohn and H. Holden, "{\it
      Solvable Models in Quantum Mechanics}", Springer--Verlag, New York
      (1988);
\rp   M. Reed and B. Simon,
      "{\it Fourier Analysis and Self-Adjointness}", Academic Press,
      Orlando (1987).
\rf{12}  A. Comtet and S. Ouvry, Phys. Lett. B {\bf 225} (1989) 272;
\rp      A. Comtet, Y. Georgelin and S. Ouvry, J. Phys. A {\bf 22} (1989)
         3917.
\rf{13}  P. Giacconi, S. Ouvry and R. Soldati, Phys. Rev. D {\bf 50} (1994)
         5358.
\rf{14}  E. Corinaldesi and F. Rafeli, Am. J. Phys. {\bf 46} (1978) 1185.

\rf{15}  A. P. Prudnikov, Yu. A. Brychkov and O. I. Marichev,
      ``{\it Integrals and series}'', Gordon and Breach Science Publishers,
      New York (1992).

\rf{16}  I. S. Gradshteyn and I. M. Ryzhik, "{\it Table of integrals series
      and products}", Academic Press, San Diego (1979).

\vfill\eject
\bye